# Benchmarking and validation of global model code for negative hydrogen ion sources

Wei Yang[1,2], Sergey N. Averkin[3,4,5], Alexander V. Khrabrov[2], Igor D. Kaganovich[2,a)] and You-Nian Wang[1,b)]

[1] *Key Laboratory of Materials Modification by Laser, Ion, and Electron Beams (Ministry of Education), School of Physics, Dalian University of Technology, Dalian 116024, China*

[2] *Princeton Plasma Physics Laboratory, Princeton University, Princeton, New Jersey 08543, USA*

[3] *Tech-X Corporation, Boulder, CO 80303, USA*

[4] *Center for Integrated Plasma Studies, University of Colorado, Boulder, CO 80305, USA*

[5] *Aerospace Engineering Program, Worcester Polytechnic Institute, Worcester, MA 01609, USA*

[a)] E-mail: ikaganov@pppl.gov

[b)] E-mail: ynwang@dlut.edu.cn

Benchmarking and validation are prerequisite for using simulation codes as predictive tools. In this work, we have developed a Global Model for Negative Hydrogen Ion Source (GMNHIS) and performed benchmarking of GMNHIS against another independently developed code, Global Enhanced Vibrational Kinetic Model (GEVKM). This is the first study to present quite comprehensive benchmarking test of this kind for models of negative hydrogen ion sources (NHIS), and very good agreement has been achieved for electron temperature, vibrational distribution function (VDF) of hydrogen molecules, and $n_{\mathrm{H}^-}/n_e$ ratio. The small discrepancies in number densities of negative hydrogen ions, positive ions, as well as hydrogen atoms can be attributed to the differences in the predicted electron number density for given discharge power. Higher electron number density obtained with GMNHIS is possibly due to fewer dissociation channels accounted for in GMNHIS, leading to smaller energy loss. In addition, we validated GMNHIS against experimental data obtained in an electron cyclotron resonance (ECR) discharge used for $\mathrm{H}^-$ production. The model qualitatively (and even quantitatively for certain conditions) reproduces the experimental $\mathrm{H}^-$ number density. The $\mathrm{H}^-$ number density as a function of pressure first increases at pressures below 12 mTorr, and then saturates for higher pressures. This dependence was analyzed by evaluating contributions from different reaction pathways to the creation and loss of the $\mathrm{H}^-$ ions. The developed codes can be used for predicting the $\mathrm{H}^-$ production, improving the performance of NHIS, and ultimately optimizing the parameters of negative ion beams for ITER.

## I. INTRODUCTION

Negative Hydrogen Ion Sources (NHIS) are used for production of energetic beams of neutral hydrogen atoms for plasma heating and plasma current drive in magnetic fusion devices, where a beam of 1 MeV energy level is required to sustain the plasma in steady state.[1] The neutralization efficiency of negative hydrogen ions remains acceptable (~60 %) for such or even higher kinetic energy, and is almost independent of the beam energy above 100 keV/nucleon. On the other hand, for positive ions the neutralization efficiency rapidly decreases as the beam energy exceeds 60 keV/nucleon, and becomes negligible in the MeV energy range.[2] Negative ions, instead of positive ions, are to be utilized for ITER and future fusion reactors because of their high neutralization efficiency at high energies required.

The production mechanisms in NHIS are classified into two types: surface production and volume production. In a surface source, energetic H atoms and positive ions are converted into $\mathrm{H}^-$ ions on surfaces covered with a low-work-function material as they collide with the chamber walls or electrodes.[3] Deposition of a material with low work function (typically cesium) on the electrode is required to enhance the negative hydrogen ion production.[4] Although the use of cesium allows negative ion sources to meet the specifications required for ITER,[5] it may cause problems in maintenance and also unstable source operation.[1] The above difficulties can be avoided with an ion source operating without cesium, which is based on a volume process of the $\mathrm{H}^-$ ions production. In such a process the $\mathrm{H}^-$ ions are formed directly in the plasma volume through the following two-step process:[6]

$$\mathrm{H}_2(\upsilon) + \mathrm{e}\ (\varepsilon > 12\ \mathrm{eV}) \rightleftharpoons \mathrm{H}_2(\upsilon') + \mathrm{e} + h\nu \quad (\mathrm{EV}), \tag{1}$$

$$\mathrm{H}_2(\upsilon) + \mathrm{e}\ (\varepsilon < 2\ \mathrm{eV}) \to \mathrm{H}^- + \mathrm{H} \quad (\mathrm{DA}). \tag{2}$$

In the first step, vibrationally-excited hydrogen molecules are produced mainly in collisions with fast electrons through excitation (EV), which refers to excitation to singlet states followed by radiative decay. In the second step electrons attach to vibrationally-excited hydrogen molecules in the process of dissociative attachment (DA) to form negative hydrogen ions. It was theoretically[7] and experimentally[6] proven that vibrationally excited hydrogen molecules have significantly higher cross section for the DA reaction. Because the DA reaction makes a major contribution to the volume production of negative ions, it was extensively studied in the past.[8-10] The cross section of the DA process is strongly dependent on both the electron energy and the initial vibrational state.[7] The vibrational kinetics is governed by excitation and deactivation of vibrational levels and is characterized by the vibrational distribution function (VDF). The VDF depends on the operational parameters and design of the ion source.

To improve the performance of the volume-production NHIS, it is important to understand the mechanisms of the volume production of $H^-$ ions and related plasma physics. Most importantly, one needs to identify the plasma conditions for enhanced production of cold electrons for DA and hot electrons for EV processes, which requires sophisticated diagnostics and detailed characterization of the ion source. In traditional volume-production sources the plasma is produced in hot-filament arcs.[11] However, there is a tendency to introduce radio-frequency (RF)[1,4,12-19] or electron cyclotron resonance (ECR)[20-23] sources. Arc sources suffer from limited lifetime of the filaments, and the evaporated material may contaminate the plasma.[24] In contrast to short-lived filament sources, RF and ECR sources allow for continuous operation, which is necessary for ITER and prospective fusion reactors. Consequently, RF and ECR–based NHIS have been widely studied by many researchers. RF plasma sources have been developed and well experienced in Garching (IPP),[1,4,12-14] Padova,[15-18] and Japan[19] in the last years. A very low pressure of 0.3 Pa is required inside the NHIS to guarantee the survivability of $H^-$ ions.[12] The plasma generated in the driver chamber enters the expansion region where a magnetic filter is added at the source periphery.[12] The magnetic field is necessary to keep hot electrons ($T_e$ > 2 eV) away from the regions where the $H^-$ ions are generated, in order to avoid collisional destruction of the negative ions. Several NHIS designs have been proposed. Recently, Averkin *et al.* proposed a high current negative hydrogen ion source (HCNHIS).[25,26] The HCNHIS consists of a high-pressure RF discharge chamber where high vibrational states of molecular hydrogen are mainly generated. The plasma and gas flow in the discharge chamber is controlled (reduced) by a series of bypass tubes and enters through a nozzle into a low-pressure negative hydrogen ion production chamber, where $H^-$ ions are produced mainly by DA process. Aleiferis *et al.* studied the effect of changing the device configuration on negative hydrogen ion production by varying the resonance location in an ECR discharge.[22]

In past decades, numerical simulations have become a valuable tool for improving the understanding of discharge physics and can provide theoretical predictions, especially in cases where diagnostic measurements are expensive or difficult to achieve in an experimental setup. Global models have been developed by many researchers to study hydrogen discharges.[27-32] A verified and validated model is required to characterize the mechanism of volume production of $H^-$ ions and to further improve the NHIS performance. Here, by verification we mean a comparison between simulation results and analytic solution, while validation is a comparison with experimental data.[33] However, analytic solutions are available only in a limited number of cases and chemically reacting plasmas in NHIS is not one of them. Therefore, comparison of different codes with each other (benchmarking) and comparison of simulation results with experimental data (validation) are both important for achieving the goal of making codes usable as predictive tools.[34-36] In this paper, we perform both benchmarking and validation. First, we present benchmarking of the GMNHIS, which was developed based on the previous study[28] against the GEVKM.[25,26] The chamber used in the models is based on the RF source developed by Gao *et al.*,[37,38] but with only the driver region considered. Gao *et al.*[37,38] focused mainly on characterizing the electron properties, while in this study we focus on the $H^-$ ion production in the NHIS. Subsequently we validate the GMNHIS against experimental measurements in an ECR discharge, where the data on negative ion production under different discharge conditions is available.[22] The plasma in the experimental reactor is sustained by a 2D network of five dipolar ECR plasma sources.[22] The negative ion production is predicted using the GMNHIS with experimentally measured electron energy distribution function (EEDF).[22]

This paper is organized as follows. The simulation model, i.e. GMNHIS, subject to benchmarking and validation is described in section 2, followed by a detailed description of the plasma chemistry reaction set for $H_2$. The benchmarking results for RF discharge obtained with the two global-model codes, GMNHIS and GEVKM, are given in section 3. A comparison between GMNHIS simulation results and experimental measurements in an ECR discharge is provided in section 4. The conclusions are given in section 5.

## II. MODEL DESCRIPTION

A global model is also called a volume-averaged model or zero-dimensional model, because it does not solve numerically for the spatial variation of plasma properties, but rather relies upon an analytical solution for the respective profiles. The power is assumed to be deposited uniformly into the plasma bulk. A Maxwellian EEDF is assumed in the GMNHIS, but the code can be easily modified to account for a non-Maxwellian EEDF in calculating the reaction rate coefficients. Ions are at the same temperature as the neutral gas and this temperature is fixed in the simulation at 600 K (although modification can be easily introduced to account for gas heating). Four fundamental conservation laws are used in the formulation of the GMNHIS: mass conservation described via a particle balance for each species except for the $H_2$ molecules; charge conservation, reduced to the quasi-neutrality condition; the $H_2$ density determined by the equation of state for total particle number density, and energy conservation law expressed through power balance. The resulting balance conditions present a system of 24 coupled nonlinear ordinary differential equations which are numerically solved. The benchmarking and validation are performed, respectively, for a cylindrical RF source[37,38] with a reactor radius *R* and height *L*, and for a cube-shaped ECR matrix source.[22]

### A. Particle balance

The particle balance for neutral species other than $\mathrm{H}_2$ molecules is given as[39]

$$n_i u_{B,i}\frac{A_{\mathrm{eff1},i}}{V}-\Gamma_j\frac{A_n}{V}+\sum R_j=0\,. \tag{3}$$

The first term shows that ions reaching the walls are recycled inside the plasma as neutrals, where $n_i$ is the volume-averaged number density of ions, and $V$ is the volume of the discharge chamber; $V=\pi R^2 L$ for a cylindrical chamber and $V=XYZ$ for a rectangular parallelepiped chamber with side lengths X, Y, and Z. For a cube three side lengths are equal and $V=L^3$ . Next, $u_{B,i}=(eT_e/m_i)^{1/2}[(1+\alpha_s)/(1+\alpha_s\gamma)]^{1/2}$ is the Bohm speed of positive ions modified due to a presence of $\mathrm{H}^-$,[40,41] where *e* is the elementary charge, $T_e$ is the electron temperature measured in eV, $m_i$ is the ion mass, and $\gamma=T_e/T_-$ , $T_-$ being the temperature of the negative ions. The electronegativity at the sheath edge $\alpha_s$ is determined based on the work of Thorsteinsson and Gudmundsson.[41] The effective areas are given for the two chamber shapes respectively as[39,42]

$$A_{\mathrm{eff1},\,i}=(2-\beta)\pi R^2 h_{L,i}/\Lambda_{L,i}+2\pi RLh_{R,i}/\Lambda_{R,i}\,, \tag{4}$$

$$A_{\mathrm{eff1},\,i}=\left(2-\beta\right)XYh_{z,i}/\Lambda_{z,i}+2YZh_{x,i}/\Lambda_{x,i}+2XZh_{y,i}/\Lambda_{y,i}\,, \tag{5}$$

where $\beta=1$ corresponds to a perfect open boundary at the bottom of the source chamber. In that case, the ions flowing out of the bottom of the source chamber do not recycle as neutrals. The value $\beta=0$ corresponds to a closed boundary. The quantity $h_i$ represents edge-to-center density ratio for the positive ion species numbered $i$. For a cylindrical chamber, it is given as[43]

$$h_{R,i}=0.80\left(4+\frac{\eta R}{\lambda_i}+\left(\frac{0.80Ru_{B,i}}{\chi_{01}J_1(\chi_{01})D_{a,i}}\right)^2\right)^{-1/2}\Bigg/\left(1+\alpha_0\right)\,, \tag{6}$$

$$h_{L,i}=0.86\left(3+\frac{\eta L}{2\lambda_i}+\left(\frac{0.86Lu_{B,i}}{\pi D_{a,i}}\right)^2\right)^{-1/2}\Bigg/\left(1+\alpha_0\right). \tag{7}$$

For a rectangular parallelepiped (box) chamber, it is given as

$$h_{L_{edge},i}=0.86\left(3+\frac{\eta L_{edge}}{2\lambda_i}+\left(\frac{0.86L_{edge}u_{B,i}}{\pi D_{a,i}}\right)^2\right)^{-1/2}\Bigg/\left(1+\alpha_0\right), \tag{8}$$

where $L_{edge}$ is *X*, *Y* or *Z*, $D_{a,i}=D_i(1+\gamma+\gamma\alpha_s)/(1+\gamma\alpha_s)$ is the ambipolar diffusion coefficient with the respective diffusion coefficient $D_i$ for positive ions, and $\lambda_i$ is the ion mean free path. $D_i$ and $\lambda_i$ will be further discussed later. $\alpha_0 \approx (3/2)\alpha$ is the central electronegativity,[43] and $\eta = 2T_+/(T_+ + T_-) = 1$ based on the assumptions of equal temperatures of ions and gas. All $\Lambda_i$ are the ratios of the volume-averaged to center densities. For cylindrical chamber, they are given as[26]

$$\Lambda_{R,i} = 0.70 b_{R,i}/(1+b_{R,i}) + 2/\chi_{01} J_1(\chi_{01})/(1+b_{R,i}), \tag{9}$$

$$\Lambda_{L,i} = 0.85 b_{L,i}/(1+b_{L,i}) + 2/(\pi + \pi b_{L,i}), \tag{10}$$

where $J_1(\chi)$ is the first-order Bessel function, and $\chi_{01}$ is the first zero of the zero-order Bessel function $J_0(\chi)$. For a rectangular parallelepiped chamber, $\Lambda_i$ is given as

$$\Lambda_{L_{edge},i} = 0.85 b_{L_{edge},i}/(1+b_{L_{edge},i}) + 2/(\pi + \pi b_{L_{edge},i}). \tag{11}$$

We have $b_{R,i} = 2(\lambda_i/R)/(T_e/T_{gas})$, $b_{L,i} = 2(\lambda_i/L)/(T_e/T_{gas})$, and $b_{L_{edge},i} = 2(\lambda_i/L_{edge})/(T_e/T_{gas})$,[26] where $T_{gas}$ is the gas temperature.

The second term in Eq. (3) accounts for the neutrals flowing out of the bottom of the chamber if the bottom is not closed, where $A_n = \beta\pi R^2$ for cylindrical chamber and $A_n = \beta XY$ for a box. $\Gamma_j = n_j v_j/4$ is the thermal flux,[39] where $n_j$ is the number density of neutrals, and the mean velocity of neutral species is $v_j = \left(8eT_j/\pi m_j\right)^{1/2}$. $m_j$ and $T_j$ are, respectively, the mass and temperature of the *j*-th species. The quantity $R_j$ in the third term of Eq. (3) is the reaction rate for the creation/loss process of neutral species *j*.

The particle balance for the ion species is given as[39]

$$-n_i u_{B,i} \frac{A_{\mathrm{eff},\,i}}{V} + \sum R_i = 0, \tag{12}$$

where $R_i$ is the volume creation/loss rate for ion species $i$. The effective areas for ions loss for the two chamber shapes in question are given as

$$A_{\mathrm{eff},\,i} = 2\pi R^2 h_{L,i}/\Lambda_{L,i} + 2\pi RL h_{R,i}/\Lambda_{R,i}, \tag{13}$$

$$A_{\mathrm{eff},\,i} = 2XY h_{z,i}/\Lambda_{z,i} + 2YZ h_{x,i}/\Lambda_{x,i} + 2XZ h_{y,i}/\Lambda_{y,i}. \tag{14}$$

The total particle number density is constrained by the given operating pressure *P*. Moreover, the discharge is assumed to satisfy the quasi-neutrality condition. These two criteria can be expressed respectively as

$$P = \sum_j n_j k_B T_{gas}, \tag{15}$$

$$n_e = n_{\mathrm{H}_3^+} + n_{\mathrm{H}_2^+} + n_{\mathrm{H}^+} - n_{\mathrm{H}^-}, \tag{16}$$

where $k_B$ is the Boltzmann constant.

## B. Power balance

The power balance refers to the balance between the absorption power and the bulk power losses caused by elastic and inelastic collisions, and also power losses due to charged species flowing to the walls:

$$P_{abs} = P_V + P_W, \tag{17}$$

where $P_{abs}$ is the absorption power per unit volume. $P_V$ is the power loss per unit volume via volumetric processes:

$$P_V = n_e \sum_j \left( \sum_i n_j \varepsilon_{inel,j}^{(i)} k_{inel,j}^{(i)} + n_j \frac{3m_e}{m_j} T_e k_{el,j} \right), \tag{18}$$

where $m_e$ is the electron mass and $n_j$ is the number density of species $j$, $k_{inel,j}^{(i)}$ is the rate coefficient of an inelastic process $i$ involving species $j$, and $\varepsilon_{inel,j}^{(i)}$ is the corresponding threshold energy. $k_{el,j}$ is the rate coefficient for electron elastic scattering on species $j$.

The power loss at the chamber wall per unit volume $P_W$ is given as

$$P_W = \sum_i n_i \left( \varepsilon_i + \varepsilon_e \right) u_{B,i} \frac{A_{\mathrm{eff}}}{V}, \tag{19}$$

where $\varepsilon_e = 2T_e$ is the mean kinetic energy per each electron lost, and $\varepsilon_i = V_p + V_s$ is the mean kinetic energy per each ion lost.[40] $V_p$ is the plasma potential,[40] and $V_s$ is the sheath potential.[41]

## C. Interactions of neutral species with the surface

The interactions of neutrals with the walls include the wall recombination of ground-state H atoms into molecules, the wall quenching of excited states of H atoms,[26] and the de-excitation of vibrationally excited hydrogen molecules.[44,45] The rate coefficient $k_{s,wall}$ is given as[46]

$$k_{s,wall} = \left[ \frac{\Lambda^2}{D_{eff,s}} + \frac{2V\left(2-\gamma_s\right)}{A v_s \gamma_s} \right]^{-1}, \tag{20}$$

where $\Lambda$ is the effective diffusion length of neutral species[47] and $A$ is the surface area of the chamber wall. The wall quenching coefficients of the $\mathrm{H}_2(\upsilon)$ molecules $\gamma_{H_2(\upsilon-\upsilon')}$ are assumed independent of the wall material. It indicates that the $\mathrm{H}_2(\upsilon)$ particles are always deexcited in collisions with the walls. The coefficients $\gamma_{H_2(\upsilon-\upsilon')}$ used in this paper are based on the vibrational distribution of $\mathrm{H_2}$ molecules reflected from the walls.[44,45] The H atoms undergo collisions with the walls to become $\mathrm{H_2}$ molecules, where the recombination coefficient $\gamma_H = 0.1$ corresponding to the stainless steel walls is adopted in the model.[48] The H (n=2, 3) atoms undergoing collisions with the walls were assumed to be deexcited to the ground states of H atoms, and the quenching coefficient $\gamma_{H(\mathrm{n})}$ is set to 1 due to the lack of data. The effective multicomponent diffusion coefficients according to Blanc's law are given by[49,50]

$$\frac{1}{D_{eff,s}} = \sum_{\substack{p=1 \\ p\neq s}}^{N_s} \frac{1}{D_{ps}} \frac{n_p}{n-n_s} . \tag{21}$$

This diffusion coefficient is different from that in the work of Huh *et al.*[31] where the Knudsen diffusion seems to be double-counted. The second term of Eq. (20) is found to have the same form as Knudsen diffusion when $\gamma_s$ is equal to 1. The Knudsen diffusion is therefore not included in Eq. (21) again. The binary diffusion coefficient assuming the same temperature of heavy species is given by[50]

$$D_{ps} = \sqrt{\frac{2\pi k_B (m_p + m_s) T_{gas}}{m_p m_s}} \frac{3}{16 n \Omega_{ps}^{(1,1)}} , \tag{22}$$

where the $\Omega_{ps}^{(1,1)} = \pi\sigma_{ps}^2 \Omega_{ps}^{(1,1)^*}$ , with a reduced collision integral $\sigma_{ps}^2 \Omega_{ps}^{(1,1)^*}$ that was tabulated in the work of Capitelli *et al.*,[51] and $n$ is the total number density of both neutral species and positive ions. The ion mean free path $\lambda_i$ mentioned earlier can be obtained from the effective multicomponent diffusion coefficients as

$$\lambda_i = \frac{1}{\sum_{p=1}^{N_i} n_p \Omega_{pi}^{(1,1)}} = \frac{16n}{3\sqrt{\frac{2\pi k_B T_{gas}}{m_i}} \sum_{p=1}^{N_i} \sqrt{1+\frac{m_i}{m_p}} \frac{n_p}{D_{pi}}} . \tag{23}$$

The diffusion coefficient $D_i$ for positive ions used to calculate the ambipolar diffusion coefficients in Eqs. (6)-(8) has the same expression as Eq. (21).

### D. Chemistry mechanism of low pressure $\mathrm{H_2}$ plasma

Table I shows the kinetic reactions considered in the model. It involves electrons, ground-state molecules $\mathrm{H_2}$ , atoms H, molecular ions $\mathrm{H_3^+}$ and $\mathrm{H_2^+}$ , atomic ions $\mathrm{H^+}$ , negative ions $\mathrm{H^-}$ , 14 vibrationally-excited molecules $\mathrm{H_2}(\upsilon = 1-14)$ , and electronically excited atoms H (n=2, 3). The reaction set mainly refers to the previous study[28] and is extended based on the benchmarking against GEVKM.

Special attention is paid to the vibrational kinetics because it can affect the $H^-$ production. Some important reactions, such as the EV process of the $H_2(\upsilon = 0)$ molecules and the vibrational-translational relaxation in collisions with molecular hydrogen (VT), were omitted in the previous study.[28] The EV process of the $H_2(\upsilon = 0)$ molecules can significantly increase the densities of high-lying vibrational states of $H_2$ molecules that are the most efficient in the $H^-$ production, due to the higher number density of the $H_2(\upsilon = 0)$ vs. other vibrational states $H_2(\upsilon = 1-14)$. The VT collisions can depopulate high vibrational states $H_2(\upsilon = 10-14)$, especially at relatively high pressures. The vibrational-translational relaxation in collisions with H atoms (Vt) and the electron detachment in collisions with vibrationally excited hydrogen molecules (EDV) are also included in GMNHIS, due to their roles played at high pressures. The importance of the $H_2(\upsilon)$ wall relaxation has been stressed by Hiskes *et al.*[44] The assumption of $H_2(\upsilon)$ molecules de-exciting only to the adjacent lower state, used in the previous paper,[28] has been replaced by de-excitation to any of lower states, same as in GEVKM. The wall quenching coefficients of $H_2(\upsilon)$ molecules $\gamma_{H_2(\upsilon-\upsilon')}$ are discussed in conjunction with Eq. (20). GMNHIS also includes various electronic excitations of hydrogen molecules (Reaction 41). Due to low number densities of excited states they are not explicitly tracked in the particle balance equations, but are added to account for energy losses in the power balance. The cross sections used here are adopted from the recently published studies.[8, 9, 26, 44, 46-48, 52-66]

TABLE I. Reaction set considered in the model.

| Reaction | Description | Ref. |
|---|---|---|
| Volume reactions | | |
| 1. $e + H_2 \rightarrow e + H_2$ | Elastic scattering | 52 |
| 2. $e + H \rightarrow e + H$ | Elastic scattering | 52 |
| 3. $e + H_2 \rightarrow 2e + H^+ + H$ | Dissociative ionization | 9 |
| 4. $e + H_2 \rightarrow 2e + H_2^+$ | Molecular ionization | 9 |
| 5. $e + H_2 \rightarrow e + H + H$ | Dissociation | 53 |
| 6. $e + H_2 \rightarrow e + H + H(n{=}2, 3)$ | Dissociation | 54 |
| 7. $e + H_2 \rightarrow H^- + H$ | Dissociative electron attachment | 9 |
| 8. $e + H \rightarrow 2e + H^+$ | Ionization | 9 |
| 9. $e + H \rightarrow e + H(n{=}2, 3)$ | Electronic excitation | 9 |
| 10. $e + H(n{=}2) \rightarrow e + H(n{=}3)$ | Electronic excitation | 9 |
| 11. $e + H(n{=}2, 3) \rightarrow 2e + H^+$ | Ionization | 9 |
| 12. $e + H_2^+ \rightarrow e + H + H^+$ | Dissociative excitation | 9 |
| 13. $e + H_2^+ \rightarrow e + H(n{=}2) + H^+$ | Dissociative excitation | 54 |
| 14. $e + H_2^+ \rightarrow 2H$ | Dissociative recombination | 55 |
| 15. $e + H_3^+ \rightarrow H_2^+ + H^-$ | Dissociative recombination | 56 |
| 16. $e + H_3^+ \rightarrow 2H + H^+ + e$ | Dissociative excitation | 54 |
| 17. $e + H_3^+ \rightarrow 3H$ | Dissociative recombination | 55 |
| 18. $e + H_3^+ \rightarrow H + H_2$ | Dissociative recombination | 55 |
| 19. $e + H^- \rightarrow H + 2e$ | Electron detachment: ED | 54 |
| 20. $H_2^+ + H \rightarrow H^+ + H_2$ | Charge exchange | 55 |
| 21. $H_2^+ + H_2 \rightarrow H_3^+ + H$ | $H_3^+$ ion formation | 57 |
| 22. $H^+ + H^- \rightarrow 2H$ | Mutual neutralization: MN | 58 |
| 23. $H^+ + H^- \rightarrow H + H(n{=}2, 3)$ | Mutual neutralization: MN | 59 |
| 24. $H_2^+ + H^- \rightarrow 3H$ | Mutual neutralization: MN | 58 |
| 25. $H_2^+ + H^- \rightarrow H_2 + H$ | Mutual neutralization: MN | 59 |
| 26. $H_3^+ + H^- \rightarrow 4H$ | Mutual neutralization: MN | 58 |
| 27. $H_3^+ + H^- \rightarrow 2H_2$ | Mutual neutralization: MN | 59 |
| 28. $H^- + H \rightarrow e + H_2$ | Associative detachment: AD | 59 |
| 29. $H(n{=}3) \rightarrow H(n{=}2) + h\nu$ | Radiative decay | 60 |
| 30. $H(n{=}2, 3) + H_2 \rightarrow H_3^+ + e$ | Quenching of H(n) by $H_2$ | 61 |
| 31. $H(n{=}2, 3) + H_2 \rightarrow 3H$ | Quenching of H(n) by $H_2$ | 61 |
| 32. $e + H_2(\upsilon) \rightarrow e + H_2(\upsilon')$ | Resonant electron-vibration excitation: eV | 62,63 |
| 33. $e + H_2(\upsilon) \rightarrow e + H_2(\upsilon') + h\nu$ | Radiative decay and excitation: EV | 8 |
| 34. $e + H_2(\upsilon) \rightarrow e + H + H$ | Dissociation via $b^3\Sigma_u^+$ : D | 64 |
| 35. $e + H_2(\upsilon) \rightarrow H + H^-$ | Dissociative electron attachment: DA | 9 |
| 36. $e + H_2(\upsilon) \rightarrow 2e + H + H^+$ | Dissociative | 9 |
| | ionization: DI | |
| 37. $e + H_2(\upsilon) \rightarrow 2e + H_2^+$ | Ionization: I | 9 |
| 38. $H + H_2(\upsilon) \rightarrow H + H_2(\upsilon')$ | Vibrational-translational relaxation: Vt | 65 |
| 39. $H_2(\omega) + H_2(\upsilon) \rightarrow H_2(\omega) + H_2(\upsilon \pm 1)$ | Vibrational-translational relaxation: VT | 59 |
| 40. $H^- + H_2(\upsilon) \rightarrow H_2(\upsilon - 2) + H + e, (2 \le \upsilon \le 6)$ | Electron detachment in collisions with $H_2(\upsilon)$: EDV | 66 |
| 41. $e + H_2 \rightarrow e + H_2(b^3\Sigma_u^+, a^3\Sigma_g^+, c^3\Pi_u, B^1\Sigma_u^+, C^1\Pi_u, E, F^1\Sigma_g^+, e^3\Sigma_u^+)$ | Electronic excitation | 53 |
| Surface reactions | | |
| 42. $H_3^+ + \text{wall} \rightarrow H_2 + H$ | Ion wall recombination | 47 |
| 43. $H_2^+ + \text{wall} \rightarrow H_2$ | Ion wall recombination | 47 |
| 44. $H^+ + \text{wall} \rightarrow H$ | Ion wall recombination | 47 |
| 45. $H + H + \text{wall} \rightarrow H_2$ | H wall recombination | 46,48 |
| 46. H (n=2, 3) + wall $\rightarrow$ H | H(n) wall recombination | 26,46 |
| 47. $H_2(\upsilon) + \text{wall} \rightarrow H_2(\upsilon'), (\upsilon' < \upsilon)$ | Vibrational de-excitation: WD | 44,46 |

## III. CODE BENCHMARKING OF GMNHIS AGAINST GEVKM FOR RF DISCHARGE

Code benchmarking of GMNHIS against GEVKM[25,26] is implemented for a cylindrical RF discharge.[38] The source chamber is a quartz tube with a diameter of 12 cm and a height of 14 cm. The absorption power is 1000 W, and the gas pressure varies from 1 to 30 Pa. Some differences between the two models must be clarified. The number density of $H_2$ molecules in GEVKM is calculated using particle balance equation with accounting for pumping of feedstock gas in and out of the discharge chamber. In GMNHIS, the equation of state is implemented to obtain the number density of $H_2$ molecules. The effect of this difference on the plasma parameters is found to be negligible in the investigated discharge parameter range. Even if the sets of volume reactions considered are not exactly the same, the reactions significantly affecting the plasma parameters are the same between these two models. Additional heavy-particle collisions included in GEVKM[26] are excluded in GMNHIS due to the relatively low range of pressures of interest in this study.

### A. $H^-$ number density and $n_{H^-}/n_e$ ratio as functions of pressure

$H^-$ number density and $n_{H^-}/n_e$ ratio obtained in simulations with two codes are shown, vs. pressure, respectively in Figures 1(a) and (b). As for $H^-$ number density, the two codes are in good qualitative and quantitative agreement in the investigated pressure range, except for the slight difference at relatively high pressures where GMNHIS predicts a higher value. The $H^-$ number density first increases and then tends to saturate with increasing pressure. Figure 1(b) shows $n_{H^-}/n_e$ ratio as a function of pressure, and an excellent agreement is achieved between the two models. In order to understand the small discrepancy in the $H^-$ number density at higher pressures, the electron properties are investigated below.

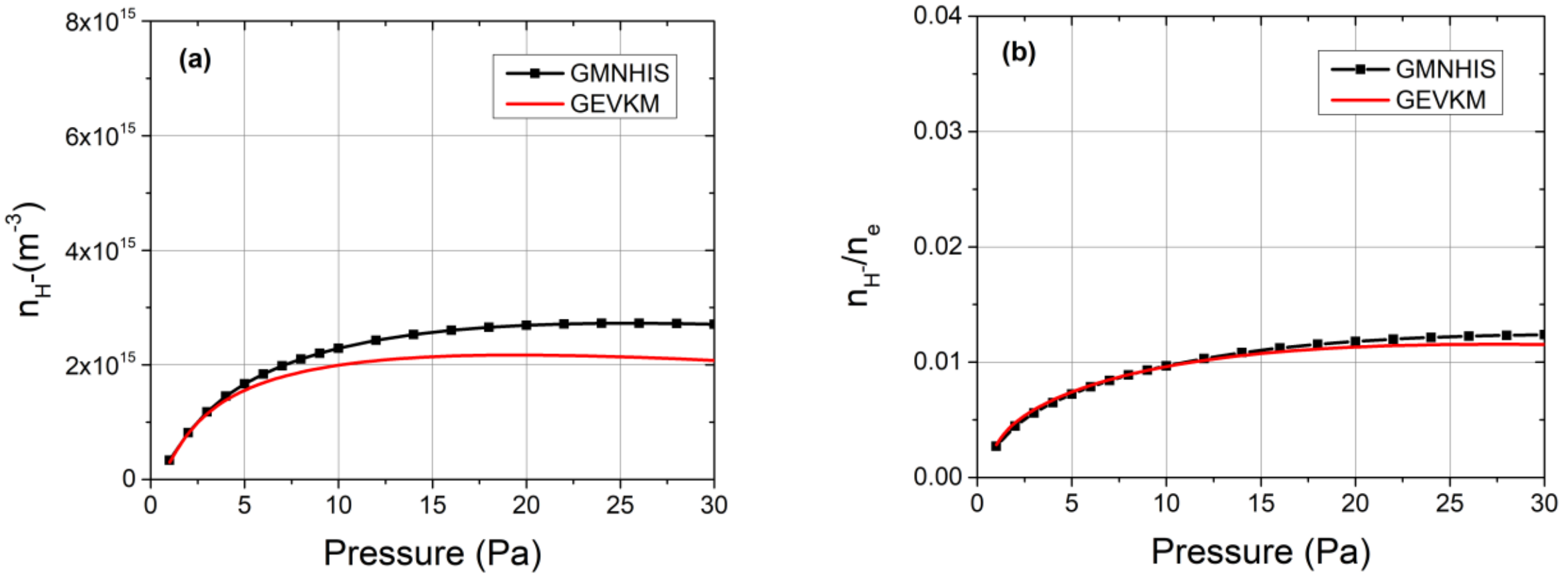

Figure 1. $H^-$ number density (a) and $n_{H^-}/n_e$ ratio (b) as functions of pressure.

## B. Electron temperature and electron number density as functions of pressure

In Figures 2(a) and (b), the electron temperature and electron number density are shown vs. pressure respectively. Excellent agreement in electron temperature is achieved between the codes in the investigated pressure range except for 1 Pa. Both codes predict the electron temperature to decrease with increasing pressure, because higher rate of elastic and inelastic collisions results in increased electron energy loss. The electron number density shown in Figure 2(b) is higher in the GMNHIS simulation than it is in the GEVKM simulation at relatively high pressures, possibly due to fewer dissociation channels considered in GMNHIS leading to a smaller energy loss in Eq. (18). The higher electron number density possibly leads to a higher $H^-$ number density in the GMNHIS simulation at relatively high pressure where most of electrons with low energy promote the production of the $H^-$ ions. In addition to electrons $H_2(\upsilon)$ molecules are the other reactants in the DA processes responsible for $H^-$ production, and the comparison of the VDFs predicted by the two models is presented in what follows.

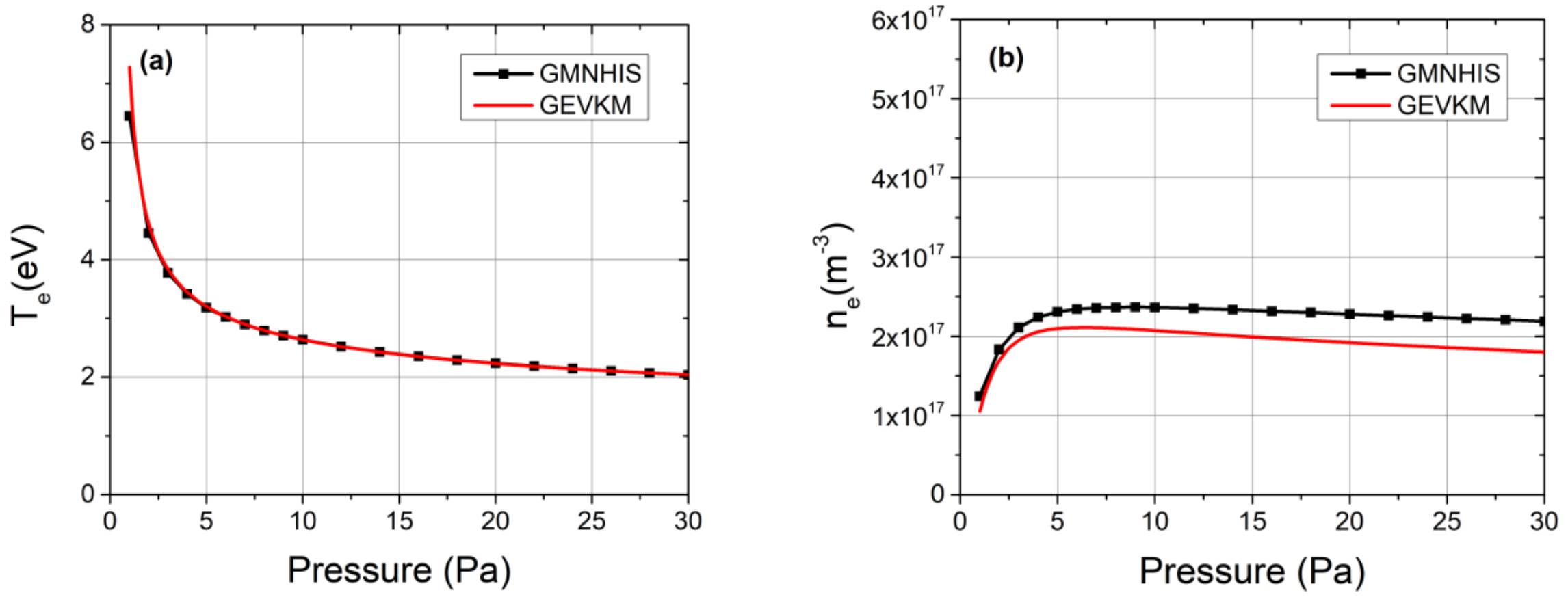

Figure 2. Electron temperature (a) and electron number density (b) as functions of pressure.

## C. VDFs at two different pressures

Figure 3 shows the VDF at two pressures: 1 Pa and 10 Pa. The density values predicted by the two codes are in very good qualitative and quantitative agreement in both cases. Both codes show the VDF is non-Boltzmann, characterized by a plateau at the intermediate vibrational levels, and the higher pressure leads to a significant decrease in the density of very high vibrational states due to the VT processes. Therefore, the higher $H^-$ number density predicted by GNNHIS is mainly due to a higher electron number density, regarding almost the same VDFs and electron temperatures predicted by the two codes.

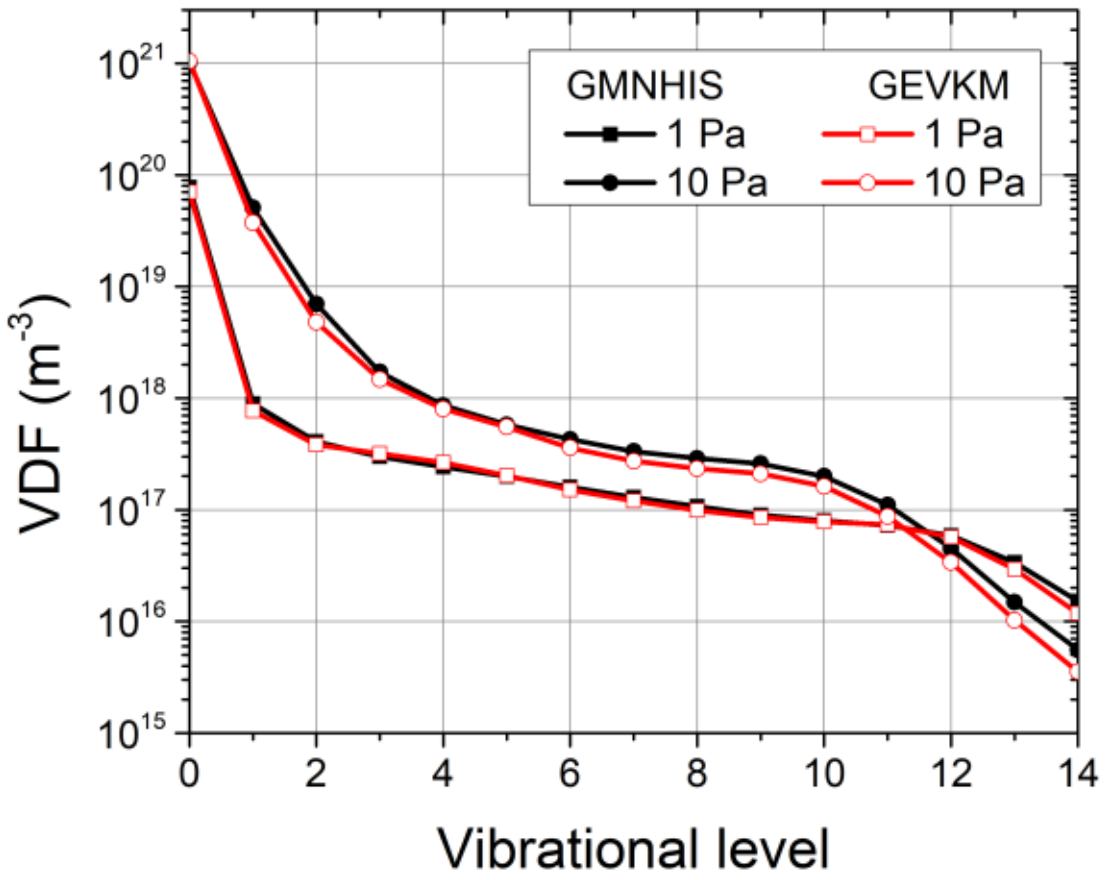

Figure 3. Predicted vibrational distributions of $H_2$ molecules at 1 Pa and at 10 Pa.

## D. Positive ion number density and H(n=1-3) atom number density as functions of pressure

Figure 4 shows the number densities of three positive-ion species as functions of pressure. The two codes are in very good qualitative agreement. As the pressure increases, $H_2^+$ number density decreases, $H^+$ number density first slightly increases and then decreases, and $H_3^+$ number density first increases and then gradually decreases. The variation in the $H_3^+$ number density versus pressure is very similar to that of electron number density, since $H_3^+$ is the prevailing ion species. The higher $H_3^+$ number density at higher pressures obtained with GMNHIS results in a higher electron number density. In this work, the most important reaction producing the $H_3^+$ is the collision between $H(n=2)$ and $H_2$, i.e., reaction 30 in Table 1. The number densities of electronically excited H atoms and the ground-state H atoms vs. pressure are shown in Figures 5(a) and (b), respectively. At higher pressures, number densities of $H(n=2, 3)$ predicted by GMNHIS are higher than those calculated by GEVKM, and thus a higher $H_3^+$ number density is found with GMNHIS. The higher production of $H(n=2)$ predicted by GMNHIS is due to higher number density of H atoms, since electronic excitation of $H$ atoms is the most important source for $H(n=2)$ at pressures above 2 Pa. The higher H atom number density in GMNHIS at higher pressures is due to high electron number density which promotes the dissociation of $H_2$. Indeed, a lower number density of $H_2$ molecules is predicted by the GMNHIS at higher pressures compared to GEVKM, although the comparison is not presented here.

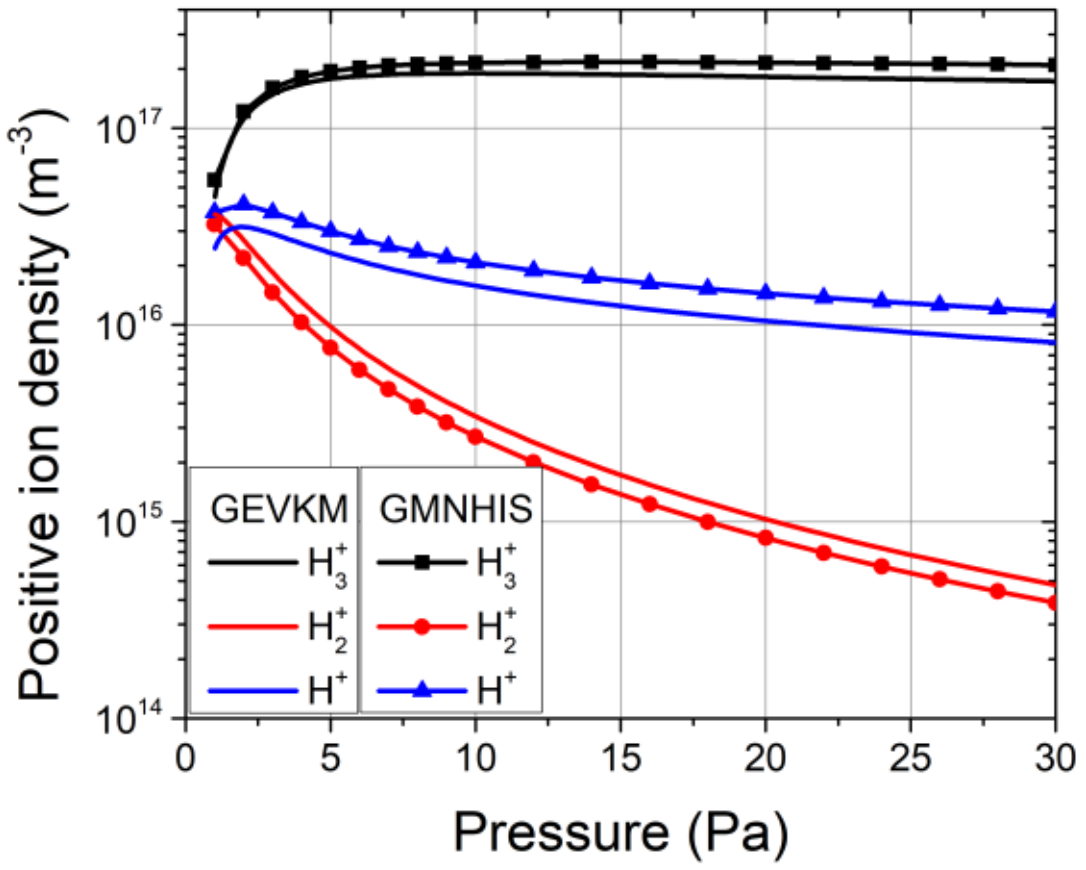

Figure 4. Positive ion number density as a function of pressure.

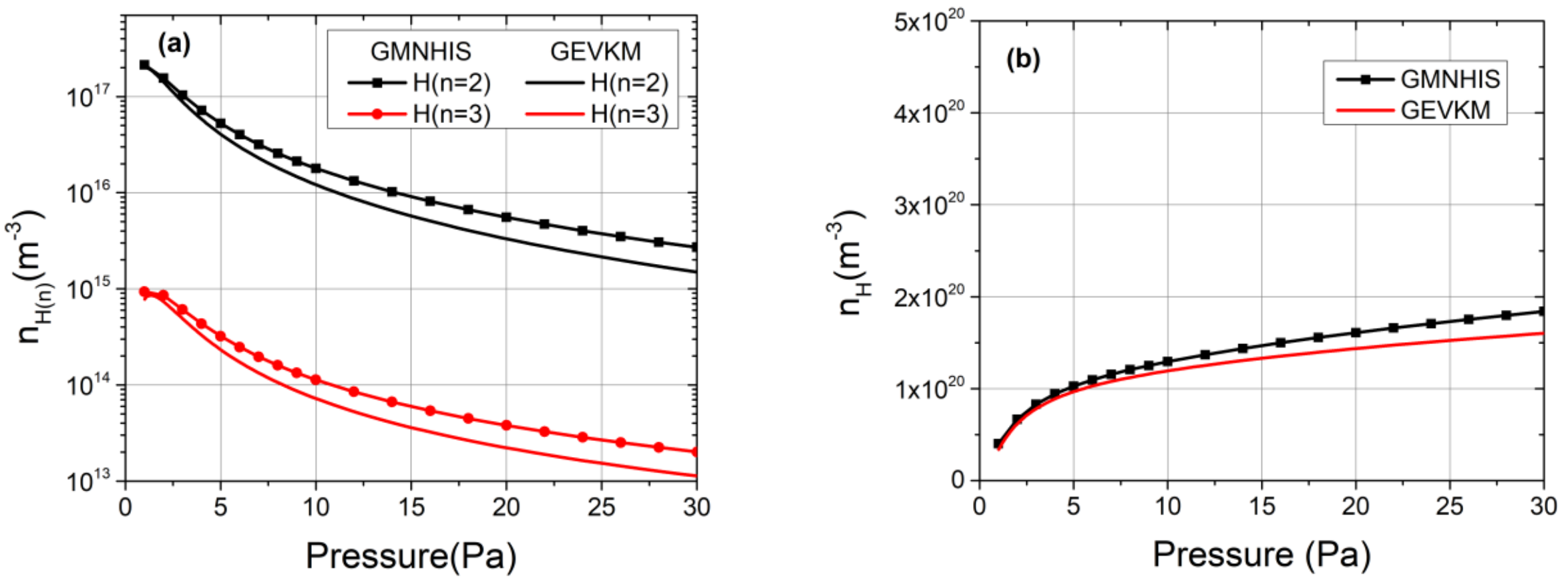

Figure 5. Number densities of electronically excited H atoms (a) and of ground-state H atoms (b) as functions of pressure.

## IV. VALIDATION OF GMNHIS WITH EXPERIMENTS FOR ECR DISCHARGE

Validation of GMNHIS outlined in Section 2 has been performed for an ECR discharge. Figure 6 shows the side view (left) and top view (right) of the experimental ECR source.[22,23] It consists of a cube-shaped stainless-steel chamber with necessary viewports for plasma diagnostics. The side lengths X, Y, and Z of the chamber, referred to in Section 2, measure 24 cm. The plasma is sustained by a matrix arrangement of five dipolar ECR plasma sources, represented by five circles in the top view of the experimental setup. The power supply of each individual source can provide up to 180 W. Such multipolar sources can produce uniform large size low-pressure plasmas. The experimental measurements are taken at the center axis 6.5 cm below the magnets. The plasma is considered to only exist in the volume below the magnets due to the presence of grounded guides above the magnets.[67] The input power is assumed to be uniformly absorbed by the plasma that extends vertically down to 6.5 cm below the magnets ("upper" region in the figure) where the discharge is concentrated. The plasma is considered to freely diffuse between the edge of the upper region and the bottom of the chamber (i.e. in the "lower" portion of the volume) where no ionization is assumed. A perfect open-boundary condition discussed in conjunction with Eq. (5) is imposed in GMNHIS for predicting the plasma parameters in the upper part of the chamber where we assume the discharge is contained. The number densities of different species at the location where the experimental measurements were taken (axially 6.5 cm below the magnets) are assumed to correspond to volume-averaged number densities obtained with GMNHIS. The absorption power is fixed at 900 W in the simulations. Four cases with different operation pressures are considered, namely 4, 8, 12, and 18 mTorr.

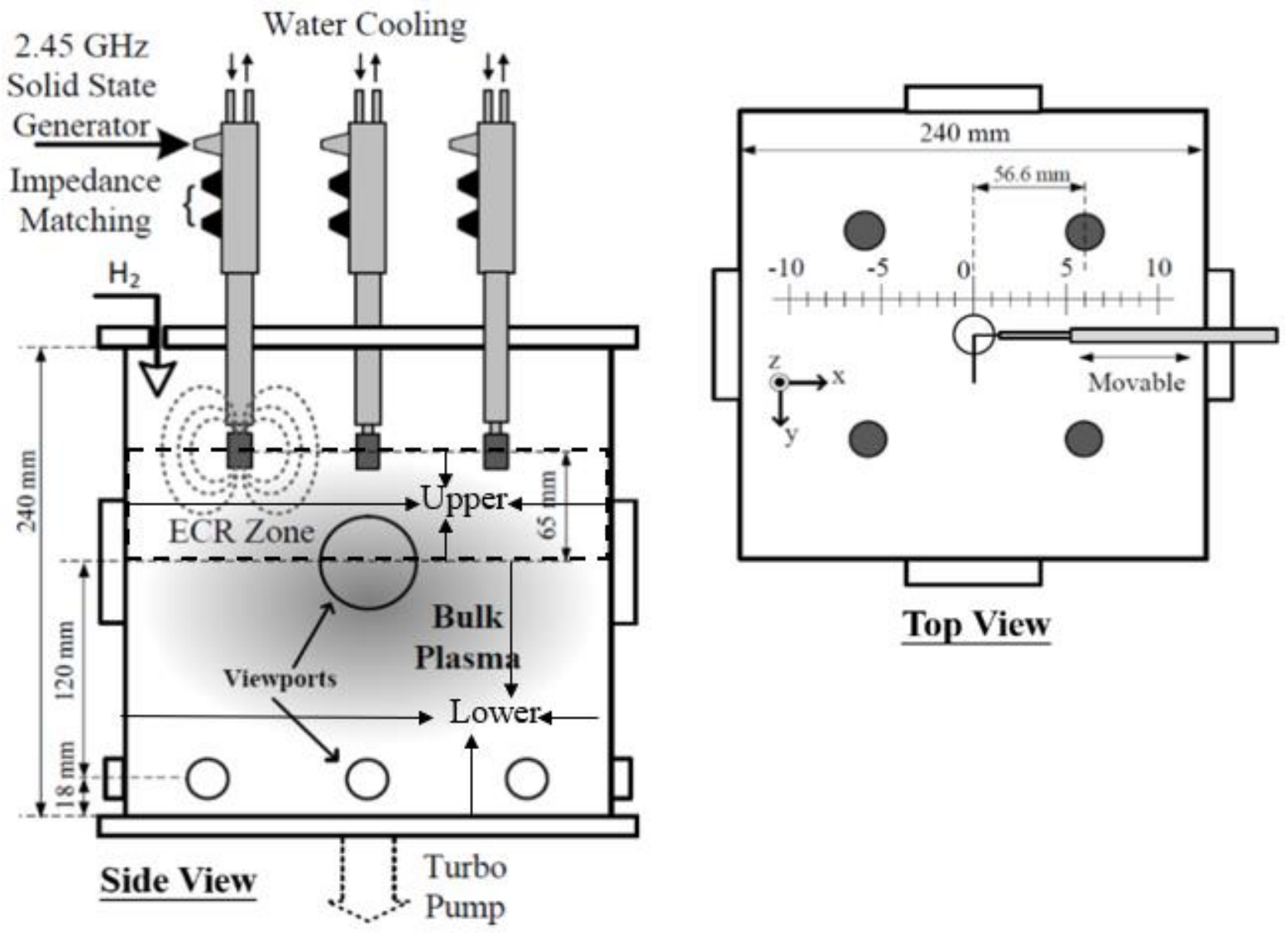

Figure 6. Side view (left) and top view (right) of experimental setup of the ECR source.[22,23]

## A. Electron number density as a function of pressure

Electron number density as a function of pressure is shown in Figure 7. The experimental data is adopted from Ref. 22. As shown in Figure 7(a), the electron number density shows first rapid increase, and then sluggish growth with increasing pressure. This dependence is qualitatively reproduced by GMNHIS. Quantitatively, electron number density predicted by GMNHIS is higher than the experimental values. It could be due to assuming all of the input power to be deposited into a small region (upper part) of the chamber. Electrons with higher energy in a smaller volume lead to a higher ionization rate, and therefore a higher electron number density is predicted by GMNHIS. Figure 7(b) shows the experimental cold and hot electron number densities as functions of pressure. The cold electron number density first increases at pressures below 12 mTorr and then saturates for higher pressures. The hot electron number density keeps increasing with the pressure.

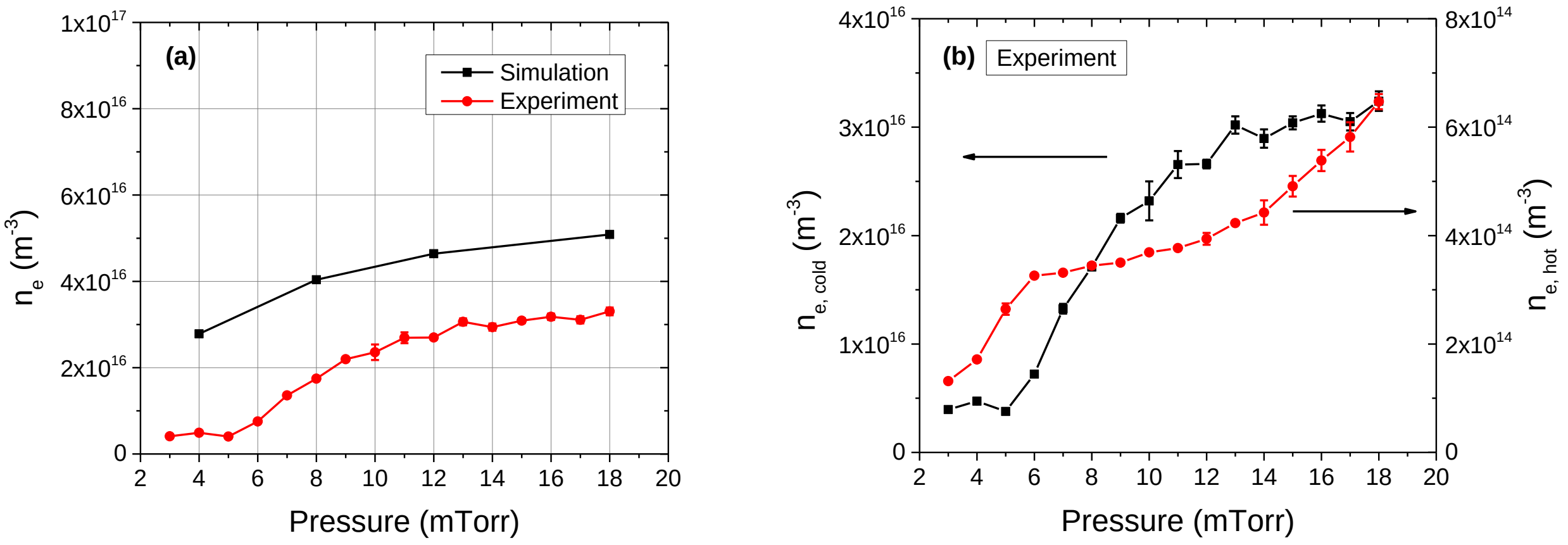

Figure 7. (a) Electron number density as a function of pressure and (b) cold and hot electron number densities as functions of pressure. The experimental data is taken from Ref. 22.

## B. Pressure dependence of VDF

The production rate of $H^-$ ions is very sensitive to the VDF, in addition to electron properties. Figure 8 shows the VDF at different pressures predicted by GMNHIS as (a) the logarithmic plot for all vibrational states $(0 \le \upsilon \le 14)$ and (b) the linear plot for higher vibrational states $(4 \le \upsilon \le 12)$. It is clear that the densities of higher vibrational states $(4 \le \upsilon \le 11)$ almost monotonically increase with increasing pressure. The significant relative decrease of $H_2(\upsilon = 12)$ density at high pressure is due to the VT processes.

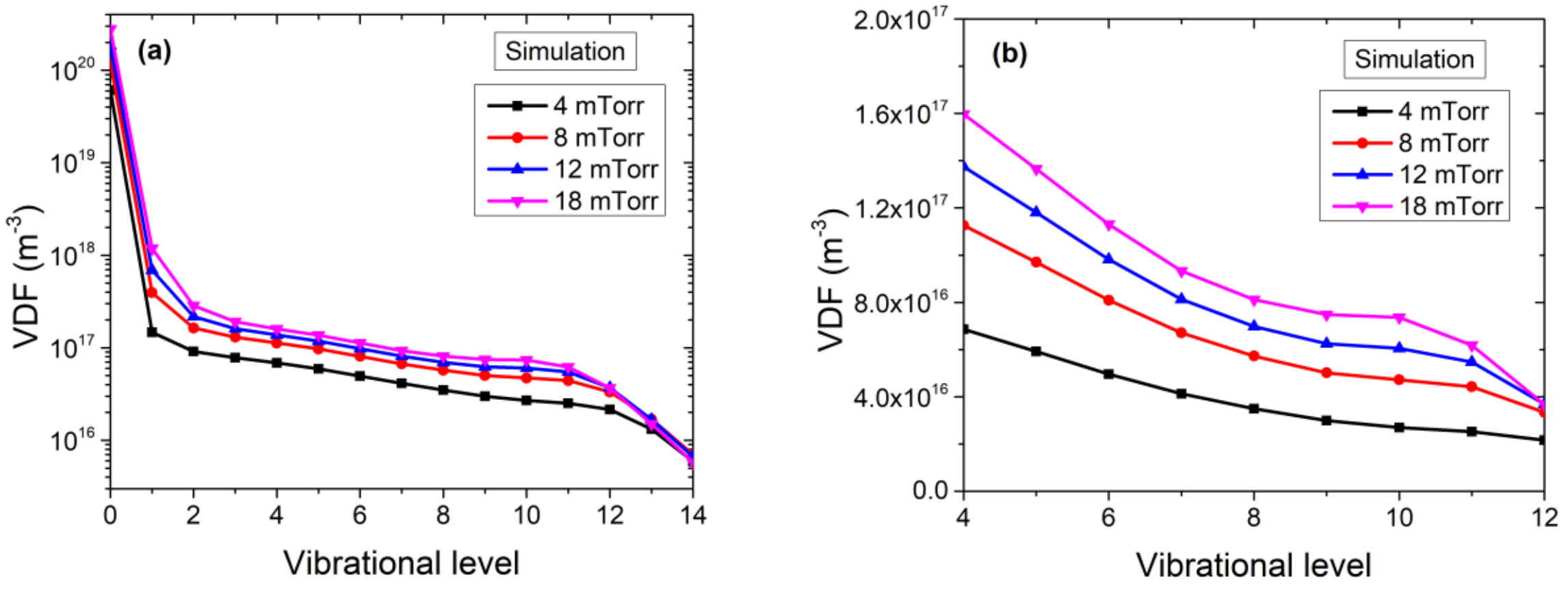

Figure 8. VDF at different pressures shown as (a) the logarithmic plot for all the vibrational states $(0 \le \upsilon \le 14)$ and (b) the linear plot for higher vibrational states $(4 \le \upsilon \le 12)$.

## C. Determination of $H^-$ number density

The number density of $H^-$ ions is estimated by the following formula:

$$n_{H^-} = \frac{\sum_{v=0}^{14} n_{H_2(v)} n_e \langle \sigma_{DA}(\varepsilon), \varepsilon \rangle}{k_{MN} n_+ + k_{AD} n_H + n_e \langle \sigma_{ED}(\varepsilon), \varepsilon \rangle + k_{EDV} n_{H_2(v)}} , \quad (24)$$

where the numerator is the $H^-$ production rate via the DA of electrons to $H_2(\upsilon)$ molecules, while the denominator includes the $H^-$ losses: mutual neutralization (MN), associative detachment (AD), electron detachment (ED), and electron detachment in collisions with vibrationally excited hydrogen molecules (EDV). The densities appearing on the right hand side of Eq. (24) are obtained using GMNHIS. The rate coefficients involving heavy species (ions and neutrals) are denoted by *k*, where the coefficients for processes involving electron collisions (DA and ED), i.e., $\langle \sigma_{DA}(\varepsilon), \varepsilon \rangle$ and $\langle \sigma_{ED}(\varepsilon), \varepsilon \rangle$, are related to the corresponding cross section $\sigma(\varepsilon)$ and the EEDF. The experimental EEDFs, approximated as bi-Maxwellians, are adopted from Ref. 22 and are shown in figure 9(a). The cross sections for DA of electrons to the selected vibrational states of $\mathrm{H_2}$ molecules are shown in figure 9(b).[9] The cross section increases by orders of magnitude with the vibrational quantum number of $\mathrm{H_2}$ molecules varying from 0 to 4. The threshold energy increases with decreasing vibrational quantum number and has a maximum value of 3.72 eV for $\upsilon=0$. Therefore, at low operating pressures considered here, low energy electrons and high vibrational states of $\mathrm{H_2}$ molecules mainly contribute to the $\mathrm{H^-}$ production.

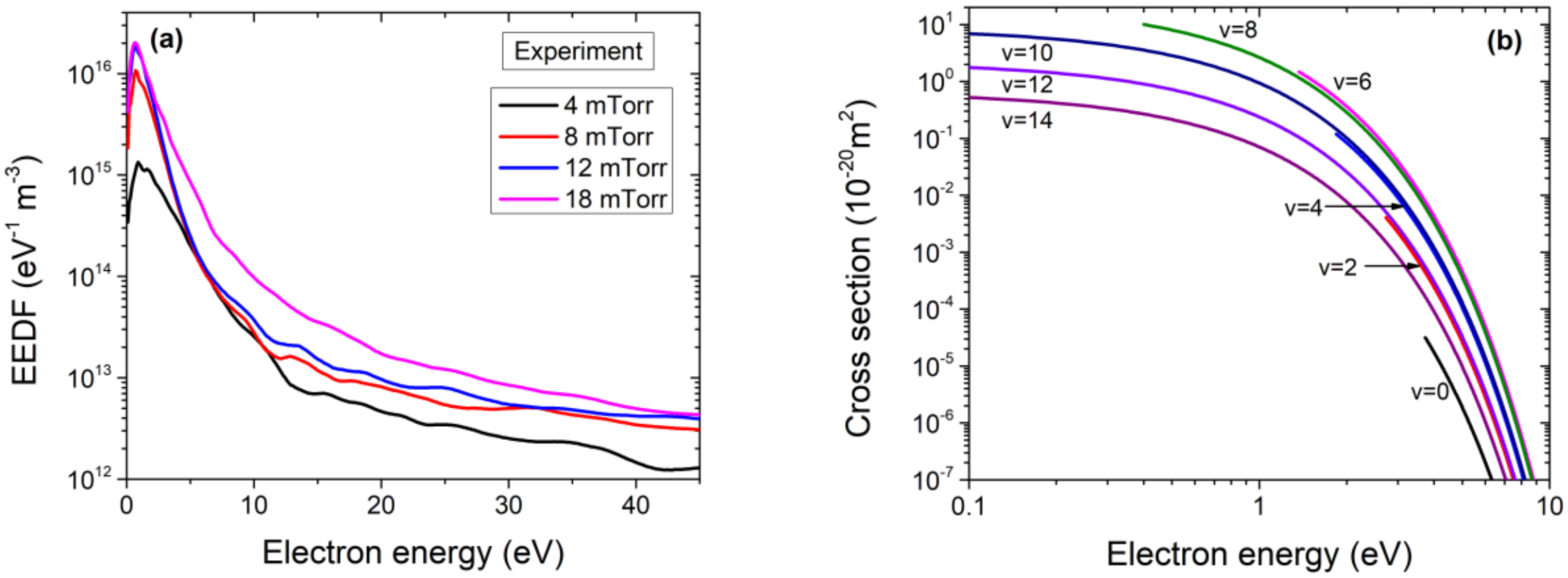

Figure 9. (a) Experimental EEDF as a function of electron energy at different pressures (Ref. 22) and (b) cross section of DA of electrons to $\mathrm{H_2}(\upsilon)$ for selected vibrational states based on the data of Ref. 9.

The experimental $\mathrm{H^-}$ number density is qualitatively reproduced by the calculation based on Eq. (24) and the comparison is shown in Figure 10(a). The $\mathrm{H^-}$ number density as a function of pressure first increases at pressures below 12 mTorr and then saturates for higher pressures. Even if the monotonous increase of the $\mathrm{H_2}(4 \le \upsilon \le 11)$ densities shown in figure 8(b) can promote the production rate of the $\mathrm{H^-}$ ions through the DA processes, the saturation of cold electron number density and the increase of hot electron number density at relatively high pressure (> 12 mTorr) can possibly limit the increase of $\mathrm{H^-}$ number density. Figure 10(b) shows the $n_{\mathrm{H^-}}/n_e$ ratio as a function of pressure. The simulation qualitatively reproduces the experimental values. The ECR source achieves a high electronegativity ($n_{\mathrm{H^-}}/n_e$) over 10 % in the pressure range from 4 to 18 mTorr. This is because the hot electrons are mostly located in the vicinity of the ECR source and do not destroy negative ions in the lower part of the chamber. The magnetic field in the upper chamber and the sheath at the wall act as filters of hot electrons and only allow the cold electrons to diffuse into the plasma bulk. Similarly, the high $n_{H^-}/n_e$ ratio achieved in the simulation is due to taking into account the experimental EEDF with large population of cold electrons that are responsible for the $H^-$ production. The $H^-$ production would be lower by an order of magnitude if the $H^-$ number density was directly calculated from GMNHIS assuming a single-temperature Maxwellian EEDF. It should also be noted that in this work we have neglected the production of $H^-$ ions due to the DA processes involving the resonant Rydberg electronic state[68] and the recombinative desorption of $H$ atoms from the walls with formation of vibrationally excited hydrogen molecules.[69] These two processes may further increase the predicted $H^-$ number density.

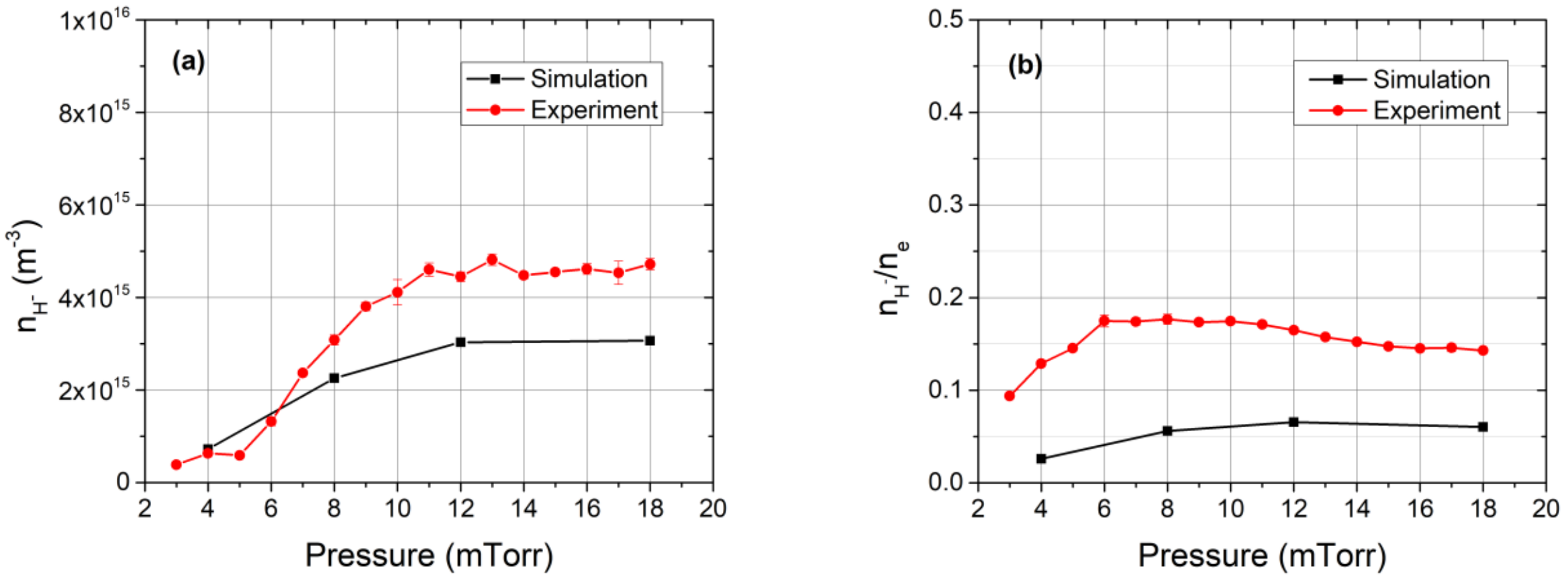

Figure 10. $H^-$ number density (a) and $n_{H^-}/n_e$ ratio (b) as functions of pressure. The experimental data is taken from Ref. 22.

## D. $H^-$ creation and loss mechanisms

In order to gain insight into the pressure dependence of $H^-$ number density, we examine the pressure dependence of the reaction rates of creation and loss of $H^-$ ions. These rates are plotted in Figure 11. The $H^-$ density is calculated from Eq. (24). As mentioned earlier, the $H^-$ ions are mainly produced by the DA of the cold electrons to high vibrational states of $H_2$ molecules. The vibrational states $H_2(\upsilon = 6-11)$ mainly contribute to the DA processes and therefore to $H^-$ production. The reaction rate of the DA processes increases with increasing pressure. At higher pressures the increase is slower due to the saturation of cold electron density. For the $H^-$ loss mechanisms, the MN processes of $H_3^+$ and $H_2^+$ with $H^-$ as well as the AD process of H with $H^-$ are the most important channels. As the pressure increases, the AD process and MN process of $H_3^+$ with $H^-$ is enhanced due to the increase in number densities of H atoms and $H_3^+$ ions, respectively. The variation of reaction rates of MN processes with pressure can be understood through presenting the number densities of $H_3^+$, $H_2^+$ and $H^+$ as functions of pressure as shown in Figure 12. The reaction rates of EDV and ED processes increase with pressure due to the increase in number densities of vibrational states and of hot electrons, but the contributions of these two processes are negligible in the investigated pressure range.

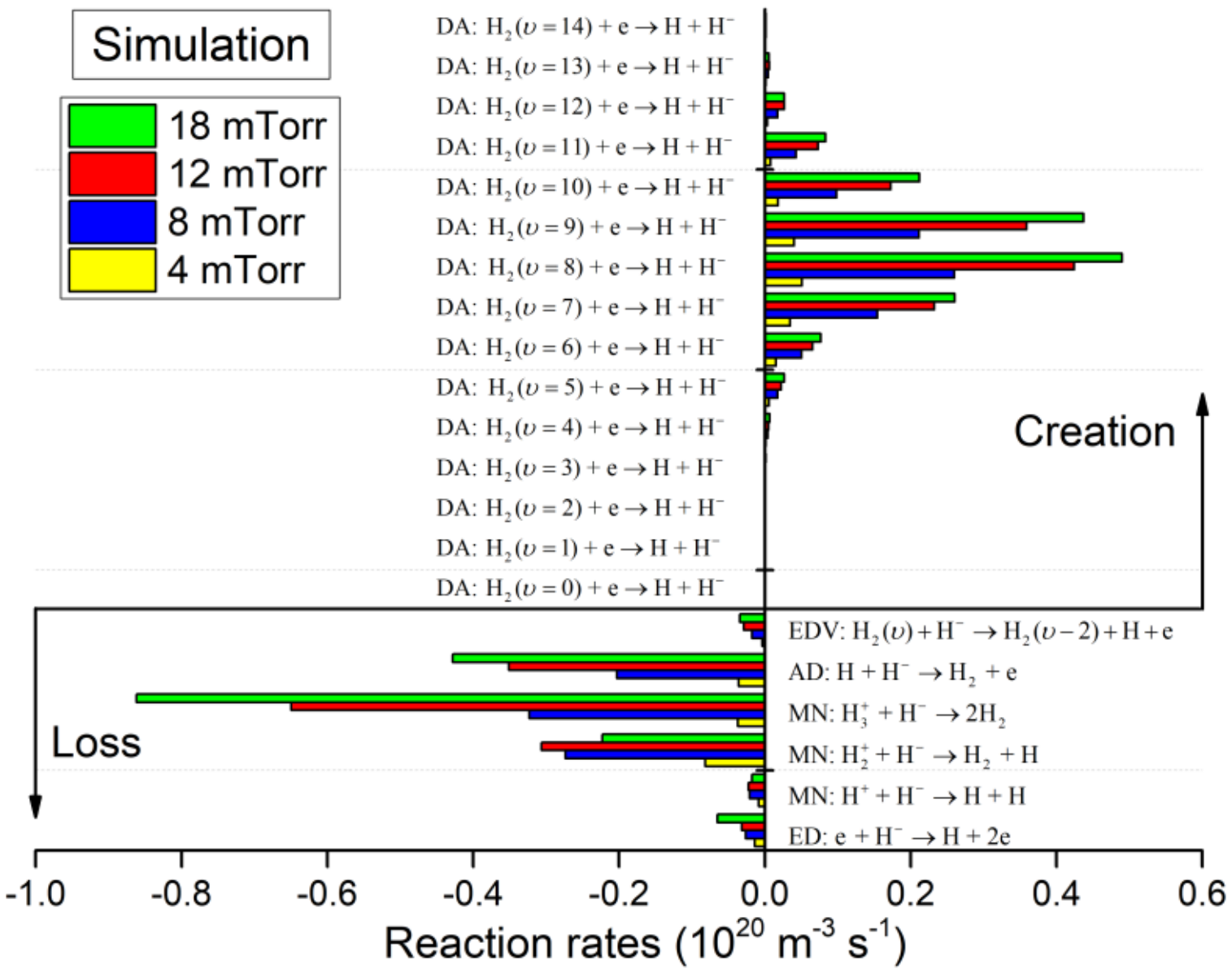

Figure 11. Reaction rates of the production and loss reactions of $H^-$ ions at different pressures.

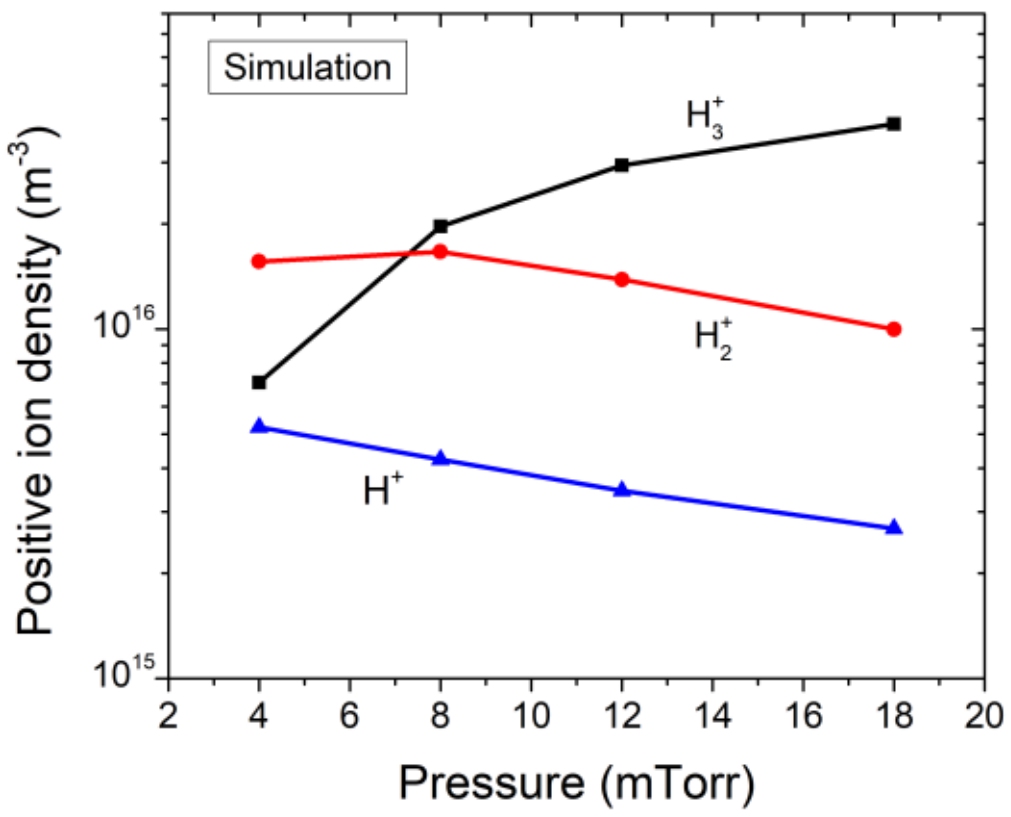

Figure 12. Positive ion number density as a function of pressure.

## V. CONCLUSION

The Global Model for Negative Hydrogen Ion Source (GMNHIS) numerical code was developed and benchmarked against an independently developed code, Global Enhanced Vibrational Kinetic Model (GEVKM). GMNHIS is based on the previous work[28] with the extended set of chemical reactions governing vibrational kinetics. Very good agreement has been achieved for pressure dependence of the plasma parameters such as vibrational distribution function (VDF), $n_{\mathrm{H}^-}/n_e$ ratio, and electron temperature. The small discrepancies observed in number densities of $\mathrm{H}^-$ ions, positive ions as well as hydrogen atoms can be due to the difference in the predicted electron number density at given discharge power. The fewer number of dissociation channels, with a smaller energy loss, considered in the GMNHIS could lead to higher electron number density predicted by GMINIS vs. GEVKM. To the authors' knowledge, this is the first study to report a comprehensive benchmarking of two numerical models of Negative Hydrogen Ion Sources (NHIS). Also, GMNHIS has been validated by comparing its predictions with experimental data obtained in measurements in an ECR discharge for $\mathrm{H}^-$ production.[22] In the simulations, the $\mathrm{H}^-$ production is predicted using GMNHIS and experimentally measured EEDF. We achieve qualitative agreement (and even quantitative agreement for certain conditions) for the $\mathrm{H}^-$ number density, which validates the simulations based on the adopted $\mathrm{H}_2$ reaction set and various assumptions used in the numerical model.

Both experiment and simulation show that the $\mathrm{H}^-$ number density first increases and then saturates with increasing pressure. In order to understand the saturation of the $\mathrm{H}^-$ production, the pressure dependence of the reaction rates for creation and loss processes of $\mathrm{H}^-$ ions has been analyzed. The specific vibrational states $\mathrm{H}_2(\upsilon = 6-11)$ mainly contribute to the dissociative attachment (DA) processes and therefore to $\mathrm{H}^-$ production. The reaction rates of the DA processes increase with pressure and at higher pressures, the increase is slower due to the saturation of cold electron density. The mutual neutralization (MN) of $\mathrm{H}_3^+$ with $\mathrm{H}^-$ and associative detachment of $\mathrm{H}$ with $\mathrm{H}^-$ (AD) mainly contribute to the $\mathrm{H}^-$ destruction, and the reaction rates of these two processes increase with pressure. Therefore, the saturation of the $\mathrm{H}^-$ production at relatively high pressures could be attributed to the saturation of cold electron number density and the enhancement in the MN process of $\mathrm{H}_3^+$ with $\mathrm{H}^-$ and the AD process.

The ECR source demonstrates good performance by producing a high $n_{\mathrm{H}^-}/n_e$ ratio over 10% in the low-pressure regime. This is because most of hot electrons are located in the vicinity of the ECR source due to the resonant heating, while cold electrons diffuse to the region away from the sources resulting in increased $H^-$ production. The possible ways to further optimize the $H^-$ production are increasing the cold electron number density by adjusting the size of upper and lower chamber or injecting a high-energy electron beam into the upper chamber, or a low-energy electron beam into the lower chamber in future $H^-$ ion beam sources. In addition, a special chamber wall material with high H atom sticking coefficient can possibly enhance the $H^-$ production by decreasing the loss of $H^-$ ions caused by the AD process. For example, tantalum and tungsten materials with a high H atom sticking coefficient of 0.5 have been proved to be $H^-$ enhancers in the work of Bentounes *et al.*,[70] according to which the production of high vibrational states of $H_2$ can be significantly increased through the interaction between surface absorbed H atoms with other H atoms.

In the future work, the DA processes involving the resonant Rydberg electronic state and the recombinative desorption of H atoms at the walls with formation of vibrationally excited hydrogen molecules will be included in our models, because these processes can increase the $H^-$ production. Benchmarking and validation studies reported here are essential to using simulation codes as reliable predictive tools, ultimately aiding in developing optimized negative ion beams for ITER and prospective fusion reactors.

## ACKNOWLEDGMENTS

This work was supported by China Scholarship Council (CSC), National Magnetic Confinement Fusion Science Program, China (Grant No. 2015GB114000) and National Key R&D Program of China (Grant No. 2017YFE0300106). The work of Igor D. Kaganovich was supported by the US Department of Energy. The authors would like to thank P. Svarnas and S. Aleiferis for providing experimental data and many helpful discussions, M. Bacal for offering helpful suggestions, and R. Celiberto for providing the cross section data for the reactions 32, 33 and 34 in Table 1.